\documentclass{aa} \usepackage{times} \usepackage{graphics}
\usepackage{xspace} \usepackage{jwaabib}

\begin{document}

\thesaurus{06(%
08.09.2 Hercules X-1; 
13.25.5; 
08.14.1; 
02.01.2)} 

\title{Evolution of spectral parameters during a pre-eclipse dip of
  Her X-1 \thanks{Table 2 is only available in electronic form at the CDS via
anonymous ftp to cdsarc.u-strasbg.fr (130.79.128.5) or via
http://cdsweb.u-strasbg.fr/Abstract.html.}}

\author{B. Stelzer\inst {1}\thanks{\emph{present address:} Max Planck
    Institut f\"ur Extraterrestrische Physik, Giessenbachstr.~1, D-85740
    Garching bei M\"unchen} \and J. Wilms\inst {1} \and R.
  Staubert\inst {1} \and D. Gruber\inst {2} \and R. Rothschild\inst
  {2}}
\institute{Institut f\"ur Astronomie und Astrophysik -- Astronomie,
  Waldh\"auser Str. 64, 72076 T\"ubingen, University of T\"ubingen,
  Germany \and
  Center for Astrophysics and Space Sciences, UCSD, La
  Jolla, CA 92093, USA}

\offprints{B. Stelzer}
\mail{B. Stelzer}
\titlerunning{Pre-eclipse dip of Her~X-1}

\date{Received $<$29 May 1998$>$ / Accepted $<$03 November 1998$>$ } 
\maketitle
 
\begin{abstract}
  We report on a pre-eclipse dip of the X-ray binary pulsar \mbox{Her X-1}
  observed by the Rossi X-ray Timing Explorer (RXTE) in 1996 July. The
  energy spectra in the 3--18\,keV range can be described by the sum of two
  power-laws, one of which is modified by photoelectric absorption and by
  Thomson scattering in cold material, plus an iron emission line at
  6.7\,keV. We present the evolution of the spectral parameters with a
  temporal resolution of 16\,s and show that the varying flux and spectrum
  can be interpreted solely by a time varying column density.  The data do
  not appear to require non-solar abundances in the absorbing material,
  although a slight over-abundance of the metals cannot be ruled out. We
  also find that the lightcurve is characterized by symmetric substructures
  which can be successfully modeled by Gaussian profiles.  The recurrence
  time of these substructures is on a timescale of a few minutes.

\keywords{stars: individual: Hercules X-1 -- X-rays: stars -- stars:
  neutron -- Accretion, accretion disks}
\end{abstract}

\section{Introduction}\label{sect:intro}
The study of transient dips in the X-ray lightcurves of X-ray binaries can
help to provide insight into the accretion processes in these objects. The
dips are thought to be caused by cold material with an appreciable column
density of $10^{23}\,\mbox{cm}^{-2}$ and higher passing through the line of
sight to the neutron star. Since the dipping activity occurs primarily
around the upper conjunction of the neutron star (i.e.,\ shortly before the
X-ray eclipse), this material is most probably due to the ``splash'' at the
location where the accretion stream hits the accretion disk. A recent
review on X-ray dip energy spectra has been published by \citey{White}.

X-ray dips of the X-ray binary \mbox{\object{Her X-1}} were first described
by \citey{Giacconi73.1}. Two different kinds of dips are seen: pre-eclipse
dips, which occur at orbital phases $\Phi_\mathrm{orb}=0.7$--$1.0$, and
anomalous dips, which are observed at $\Phi_\mathrm{orb} \sim 0.2$. During
the ``Main High'' phase of the 35\,d cycle of \mbox{Her X-1}, the pre-eclipse
dips seem to ``march'' towards earlier orbital phases
(\cite{Crosa80,Giacconi73.1}).  Interpretations of the marching behavior
have been given in terms of a periodic mass transfer driven by the changing
radiation pressure at the inner Lagrangian point (\cite{Crosa80}), and by
the splash of the accretion stream onto a warped accretion disk
(\cite{Schandl96.1}). Note that the observational basis for the ``marching
behavior'' has recently been challenged by \citey{Leahy97.1}.

Previous observational studies dedicated to pre-eclipse dips of \mbox{Her
  X-1} have been presented by \citey{Ushimaru89}, \citey{Choi94.1},
\citey{Leahy94.1}, and \citey{Reynolds95.1}. These authors found that the
dip spectra can be modeled by a temporally variable absorption of the
non-dip spectrum. In addition to this absorption component, a further weak
unabsorbed component is present during the dips, as first seen in
\textsl{Tenma} data (\cite{Ushimaru89}) and verified by later observations.
This component has been interpreted as being due to Compton scattering of
the primary neutron star radiation by an extended corona into our line of
sight.  Although the unabsorbed component always contributes to the
observed X-ray spectrum, it can only be identified during dips and during
the ``low state'' of \mbox{Her X-1} (\cite{Mihara91.1}). Due to the
comparably small effective areas of the earlier instruments, a study of the
detailed temporal evolution of the spectral parameters during a dip has not
been possible in these observations. Therefore, earlier investigators were
forced to combine data obtained during (non-consecutive) time intervals
with similar measured count-rates to obtain spectra with a signal to
noise ratio sufficient for spectral analysis. This approach is not without
problems, however, since a strong variation of $N_\mathrm{H}$ during the
time intervals used for accumulating the spectra can lead to spurious
spectral features like an apparent decrease in the flux incident onto the
absorbing material (\cite{Parmar86.1}).

To be able to resolve the structures occuring on short time\-scales,
instruments with a larger effective area and moderate energy resolution are
needed.  In this paper, therefore, we present results from the analysis of
an RXTE observation of \mbox{Her X-1}, focusing on the spectral analysis
and the behavior of the light\-curve during a pre-eclipse dip.  Results from
a previous analysis of these data, focusing on the orbit determination,
have been presented by \citey{Stelzer97.1}. In
Sect.~\ref{sec:datareduction} we describe our RXTE observation and the data
reduction. Sect.~\ref{sec:timeevol} presents our results on the temporal
evolution of the column density of the absorbing material throughout the
dip, using two different data analysis methods. In Sect.~\ref{sec:lightcurves}
we discuss the temporal behavior of the source in terms of a simple model
for individual structures in the lightcurve. We summarize our results in
Sect.~\ref{sec:conclusions}.

\section{Observations and data analysis}\label{sec:datareduction}
Our RXTE observations of \mbox{Her X-1} were performed on 1996 July~26. We
primarily use data obtained with the PCA, the low energy instrument onboard
RXTE. The PCA consists of five nearly identical Xe proportional counters
with a total effective area of about $6500\,\mbox{cm}^2$
(\cite{Jahoda96.1}).  The data extraction has been performed using the RXTE
standard data analysis software, ftools~4.0.  To avoid contamination of the
spectra due to the Earth's X-ray bright limb, only data measured at source
elevations more than $10\degr$ above the spacecraft horizon were used in
the present analysis. Background subtraction was performed with version~1.5
of the background estimator program, using a model where the background is
estimated from the rate of Very Large Events in the detector. The diffuse
X-ray background and a model for the activation of radioactivity within the
detector is added. See \citey{Jahoda96.2} for a description of the PCA.

For the spectral analysis, version 2.2.1 of the PCA response matrix was
used (Jahoda, 1997, priv.\ comm.).  The spectral modeling of data from an
observation of the Crab pulsar made available to us by the RXTE PCA and
HEXTE teams with this matrix suggests that the matrix is well understood on
the 1\% level (see \cite{Dove98.1} for a discussion of these data). Using
this version of the PCA response matrix, the overall description of the
Crab continuum is good, with the remaining deviations mainly being below
3\,keV and around the Xe K edge at 35\,keV.  To avoid these uncertainties
we used data from 3 to 18\,keV only, which is sufficient for an analysis of
the dips because of the power law nature of the Her~X-1 X-ray spectrum and
due to the $E^{-3}$ dependency of the photoionization cross section (see
discussion in Sect.~\ref{subsec:specfits}).  To obtain meaningful $\chi^2$
values and to take into account the remaining uncertainties of the matrix,
we added a 1\% systematic error to all PCA channels. The spectral analysis
was performed with XSPEC, version~10.00p (\cite{Arnaud96.1}).

\section{Time evolution of spectral parameters}\label{sec:timeevol}
\subsection{Dip lightcurves}\label{subsec:diplc}
\begin{figure}
  \resizebox{\hsize}{!}{\includegraphics{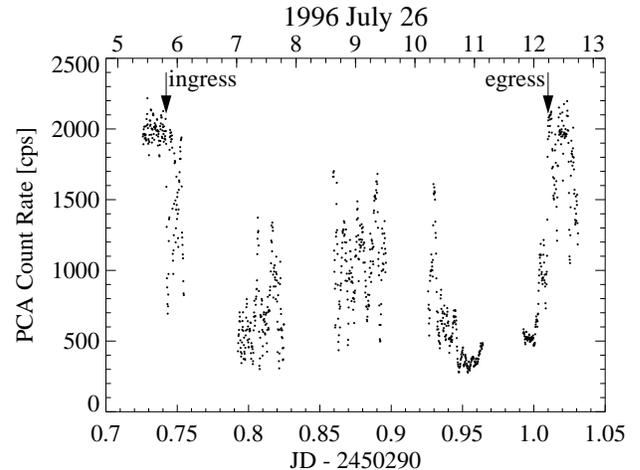}}
  \caption{Lightcurve during the dip of the RXTE observation in July
  1996, 16\,s resultion. Top $x$-axis: Time [UT]}\label{fig:diplc}
\end{figure}

The full lightcurve of the July 26 pre-eclipse dip is shown in 
Fig.~\ref{fig:diplc}. Gaps in the data stream are due to passages through
the South Atlantic Anomaly and due to the source being below the spacecraft
horizon.
The dip ingress, which occurred at orbital phase $\Phi_\mathrm{orb}=0.75$,
is characterized by a rapid decrease in intensity by a factor 3 within
80\,s.  The RXTE observation corresponds to a 35\,d phase of $\Psi_{35} \sim
0.13$, that is the observation took place 4 to 5 days after the Turn-On of
the Main High State.  From the behavior of the lightcurve in
Fig.~\ref{fig:diplc} we estimate that the dip egress takes place close to
the end of the observation, when the count rate again reaches the pre-dip
level.  Under this assumption the duration of the dip was 6.5\,hours.

In order to describe the time evolution of spectral parameters we use two
complementary methods: First, we divide the whole dip lightcurve into
segments of 16\,s duration and perform spectral fits to each of these
segments (Sect.~\ref{subsec:specfits}).  Secondly, we use color-color
diagrams to model the time evolution of the column density
(Sect.~\ref{subsec:colordiag}).

\subsection{Spectral modeling}\label{subsec:specfits}
As was mentioned in Sect.~\ref{sect:intro}, the common explanation for the
dips is that of photoabsorption and scattering by foreground material. The
photon spectrum ($\rm ph\,cm^{-2}\,s^{-1}\,keV^{-1}$) resulting from this
process can be well described by a partial covering model of the form
\begin{eqnarray}
N_\mathrm{ph} & = & E^{-\alpha}\cdot
     \left[
       I_\mathrm{A} \exp\left(-\left(
         \sigma_\mathrm{T} k N_\mathrm{H}
         + \sigma_\mathrm{bf}(E) N_\mathrm{H}
       \right)\right)
       + I_\mathrm{U}
     \right] \nonumber \\
  & & + \mbox{GAUSS}
\label{eq:pcovmod}
\end{eqnarray}
where $\sigma_\mathrm{bf}(E)$ is the photoabsorption cross section per
hydrogen atom (\cite{Morrison83.1}), $\sigma_\mathrm{T}$ is the Thomson
cross section, and $k=N_\mathrm{e}/N_\mathrm{H}$ is the number of electrons
per hydrogen atom ($N_\mathrm{e}$ is the electron column density). In
Eq.~(\ref{eq:pcovmod}), the continuum emission is modeled as the sum of two
power laws, one of which is photo-absorbed by cold matter of column density
$N_\mathrm{H}$, as well as Thomson scattered out of the line of sight by
electrons in the cold material. The second power law is not modified by the
absorber, indicating that this additional (scattering) component comes from a 
geometrically much larger, extended region and thus is not affected by the
photoabsorption. To this continuum, an iron emission line (described by a 
Gaussian) is added, which remains unabsorbed to simplify the spectral 
fitting process.

\begin{figure}
  \resizebox{\hsize}{!}{\includegraphics{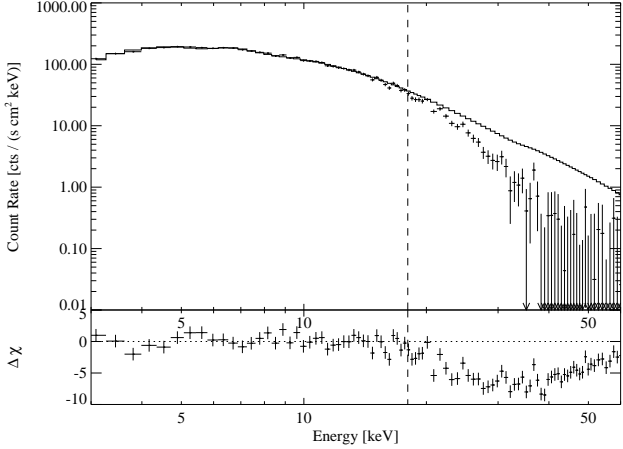}}
  \caption{16\,s PCA spectrum taken outside of the dip, modeled with the
    spectral shape from Eq.~\ref{eq:pcovmod} over the energy interval used
    for our analysis of the dip-spectra. Significant deviations from the
    power-law continuum model appear only at energies above 18\,keV,
    indicating that a power-law is sufficient in describing the data below
    18\,keV.}
  \label{fig:nondipfit}
\end{figure}

Two-component models similar to that of Eq.~(\ref{eq:pcovmod}) have
previously been shown to yield a good description of the dip spectra, while
other simpler models were found to result in unphysical spectral
parameters. 
As we mentioned in Sect.~\ref{sec:datareduction}, to avoid
response matrix uncertainties and problems with the exponential cutoff and
the cyclotron resonance feature we include only data from 3 to 18\,keV in
our analysis. This approach results in a simpler spectral model than that
used by \citey{Choi94.1} and \citey{Leahy94.1}, who included the whole
\textsl{Ginga} LAC energy band from 2 to 37\,keV in their analysis. Since
$\sigma_{\rm bf} \propto E^{-3}$, except for the highest values of $N_{\rm
  H}$, photoabsorption will virtually not influence the spectrum above
$\sim$10\,keV, such that the inclusion of data measured up to 18\,keV is
sufficient for the determination of the continuum strength. As we show in
Fig.~\ref{fig:nondipfit}, neither the exponential cut-off nor the cyclotron
resonance feature at $\sim$35\,keV need to be taken into account when
restricting the upper energy threshold to 18\,keV, as significant
deviations between the data and the model are observed only for energies
above 18\,keV. We, therefore, conclude that between 3 and 18\,keV,
additional continuum components do not affect the spectrum.  
When holding the power law index constant at its pre-dip value, from our
spectral fits of the dip data we obtain acceptable $\chi^2$ values
($\chi^2_{\rm red} \la 1.5$). Introducing additional freedoms by leaving
the photon index as a free parameter in the fit does not significantly
improve the results. Therefore, we do not have to include high energy
data to determine the continuum parameters.

\begin{figure}
  \resizebox{\hsize}{!}{\includegraphics{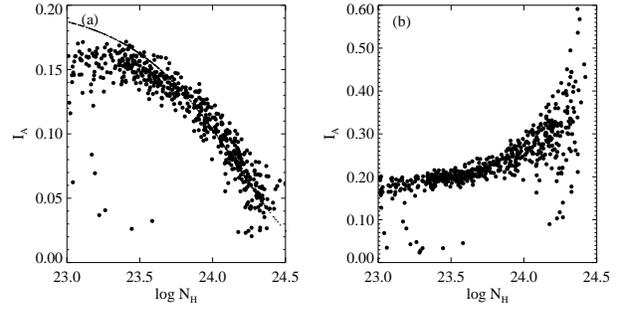}}
  \caption{\textbf{a} and \textbf{b}: Dependency of the absorbed intensity
    $I_A$ on the column density $N_{\rm H}$ for variations of the two
    component model: \textbf{a} model without inclusion of Thomson
    scattering, and \textbf{b} model with free $I_A$.  In subfigure (a) the
    line represents the exponential Thomson factor $I_A
    \exp{(-\sigma_{\rm T} N_{\rm H})}$ for $I_A=0.2\,{\rm
    ph\,cm^{-2}\,s^{-1}\,keV^{-1}}$. Uncertainties have been omitted 
    for clarity.
}
  \label{fig:Ia_nh}
\end{figure}

In different attempts to fit the 16\,s time resolved spectra without
explicitly allowing for Thomson scattering of the absorbed component we
found that the absorbed intensity which was then a free parameter
was strongly anticorrelated with the column density. The relation between
$I_\mathrm{A}$ and $N_{\rm H}$ from a fit of a spectral model 
that does not take account of Thomson
scattering is shown in Fig.~\ref{fig:Ia_nh}\,a. The observed
anticorrelation reflects the exponential $N_{\rm H}$-dependence of the Thomson
scattering factor and led us to the conclusion that absorption and Thomson
scattering may not be separated.
We, therefore, use the model of Eq.~(\ref{eq:pcovmod}) and hold
the continuum parameters fixed to their measured pre-dip values: a single
power law of photon index 1.06 plus an emission line feature from ionized
iron at 6.7\,keV with width of $\sigma=0.46$\,keV
($\chi^2_\mathrm{red}=1.1$ at $3$--$18$\,keV for 30 degrees of
freedom).  The normalization of the absorbed power law, $I_\mathrm{A}$, was
also fixed to its pre-dip value, $I_\mathrm{A}=0.20\,
\mbox{ph}\,\mbox{cm}^{-2}\,\mbox{s}^{-1}\,\mbox{keV}^{-1}$. Finally, the
ratio $k=N_\mathrm{e}/N_\mathrm{H}$ was set to $1.21$, appropriate for
material of solar abundances.  
We emphasize that fixing the parameters to
their normal state values assumes that the intrinsic spectral shape of the
source does not change during the dip and that variations of the observed
spectrum are due to the varying column density only. This is justified by
the apparent constancy of the lightcurve outside the dip.
Particularly, if $I_{\rm A}$ is free in the fit, we observe an increase of this
parameter for very high $N_{\rm H}$, which is not clearly systematic
 (see Fig.~\ref{fig:Ia_nh}\,b).
 Such a correlation between $I_{\rm A}$ and $N_{\rm H}$ seems to indicate an
 additional dependence of $I_{\rm A}$ 
 on the column density (next to the Thomson
 scattering already taken into consideration in the spectral model).
 Rather than being due to real variations of the absorbed intensity, we
 consider this relation to be produced artificially by the fitting process:
 Variations of $I_{\rm A}$ during the dip might come about as compensation for slight
 misplacements of the column density.
Any remaining correlation in Fig.~\ref{fig:Ia_nh}\,b might contain a possible
contribution from slight variations in the absorbed continuum. Due
to the limited energy resolution of the detector, however, $N_{\rm H}$
and $\alpha$ are strongly correlated. Therefore, a slight real variation
is not convincingly separable from the artificial one.

To summarize, the remaining free parameters of the spectral model
are the iron line normalization, $N_{\rm Fe}$, the normalization 
of the unabsorbed component, $I_\mathrm{U}$, and the column density,
$N_\mathrm{H}$.

\begin{figure*}
  \resizebox{\hsize}{!}{\includegraphics{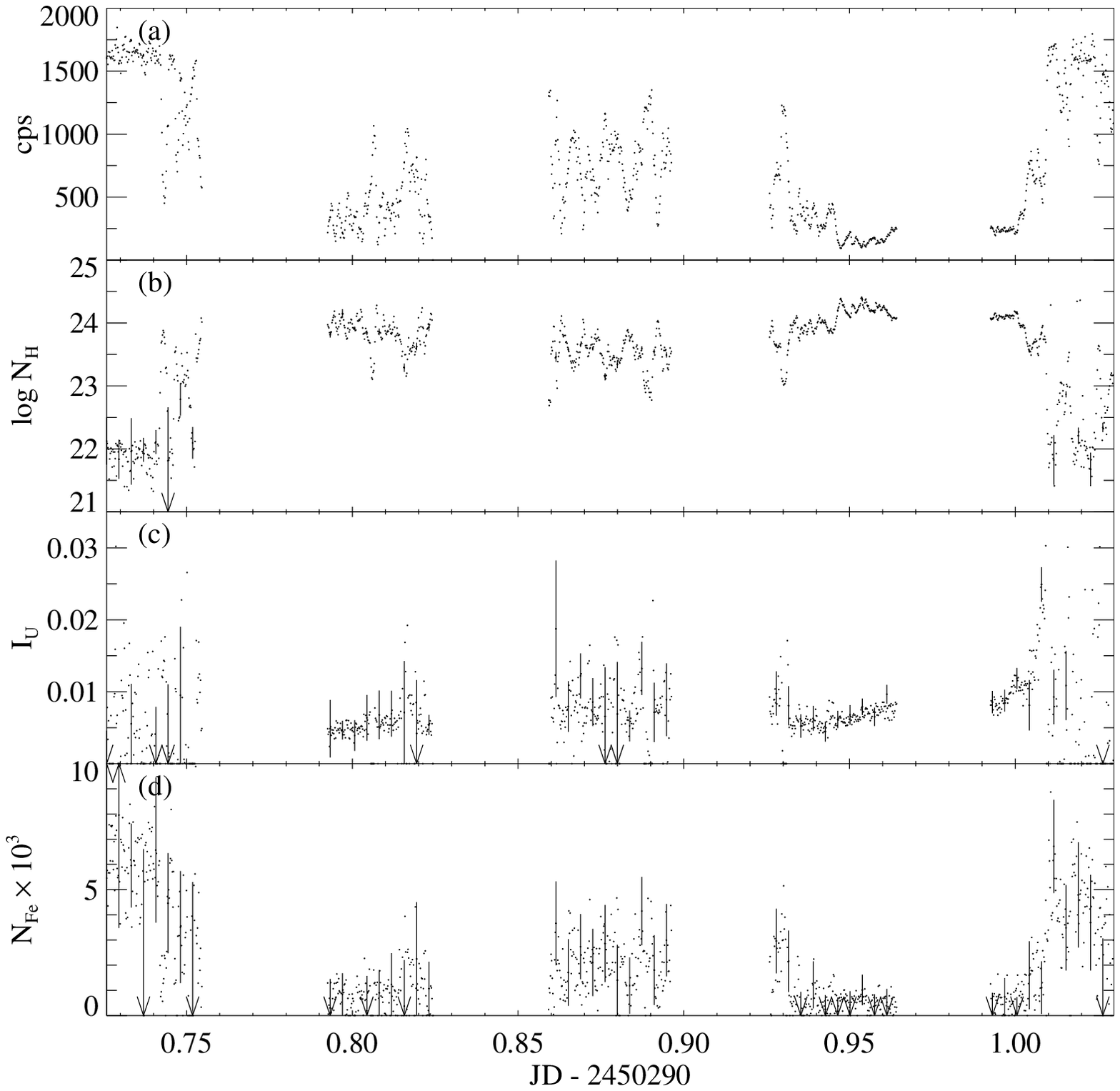}}
  \caption{\textbf{a}--\textbf{d} Time evolution of bestfit parameters from
    spectral fitting in 16\,s intervals: \textbf{a} PCA count rate for the
    energy interval from 3.5 to 17\,keV, \textbf{b} $\log N_\mathrm{H}$,
    where $N_\mathrm{H}$ is measured in $\mbox{cm}^{-2}$, \textbf{c}
    normalization of the unabsorbed power law, $I_\mathrm{U}$, in
    $\mbox{ph}\,\mbox{cm}^{-2}\,\mbox{s}^{-1}\,\mbox{keV}^{-1}$ at 1\,keV,
    \textbf{d} normalization of the Gaussian emission
    line.
    Uncertainties shown are at the 90\% level for one parameter. For
    clarity, the error bars are only shown for every 20th data point.}
\label{fig:timeevol}
\end{figure*}

\begin{figure}
  \resizebox{\hsize}{!}{\includegraphics{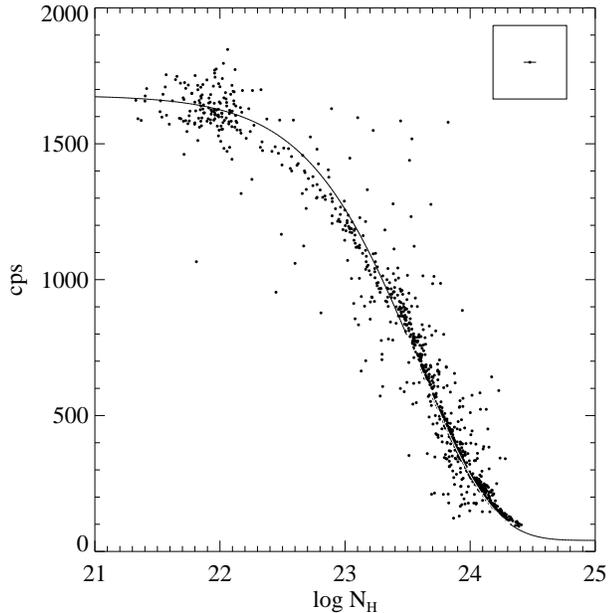}}
  \caption{PCA count rate in the band from 3 to 18\,keV as a function of
    the measured column density. The line represents the count rates
    predicted by the model of Eq.~(\ref{eq:pcovmod}). Inset: typical
    $1\,\sigma$ uncertainty for $N_\mathrm{H}$.}\label{fig:cps_nh}
\end{figure}

We divide the dip observation into 16\,s intervals and obtain 941 spectra,
covering the energy range from 3 to 18\,keV.  After subtracting the
background, which has been modeled on the same 16\,s basis, the individual
spectra were fitted with the model of Eq.~(\ref{eq:pcovmod}). Typical
$\chi^2_\mathrm{red}$ values obtained from these fits are between
$0.5$--$1.5$ for 37~degrees of freedom, indicating that our simple spectral
model is sufficient to describe the data.

The temporal behavior of the column density mirrors that of the
light\-curve (Figs.~\ref{fig:timeevol}a and b). This supports the
assumption that the underlying cause for both is absorbing material whose
presence in the line of sight blocks off the X-ray source and thus leads to
a modification of the spectral shape due to energy dependent absorption and
energy independent scattering as well as a corresponding reduction in the 3
to 18\,keV flux.  The highest value measured for the column density is
about $3\,10^{24}\,\mbox{cm}^{-2}$, comparable to that found in previous
measurements (\cite{Reynolds95.1}).  The variation of the measured count
rate with $N_\mathrm{H}$ is shown in Fig.~\ref{fig:cps_nh} together with
the count rates predicted by our partial covering model.  The transition
from the dip to the normal state is manifested in the break of the slope
around 1500\,cps. The figure indicates that due to the 3\,keV energy
threshold of the PCA, the instrument is sensitive only to values of
$N_\mathrm{H}$ of about $5\,10^{22}\,\mbox{cm}^{-2}$ and above. This also
explains why we measure $N_\mathrm{H} \simeq 10^{22}\,\mbox{cm}^{-2}$ for
the out-of-dip data just before dip-ingress, a value which is rather high
compared to the value for absorption by the interstellar medium along the
line of sight ($1.7\,10^{19}\,\mbox{cm}^{-2}$, \cite{Mavromatakis93}).

The normalization of the unabsorbed component, $I_\mathrm{U}$, is found to stay
almost constant during the whole dip, indicating that the whole variability
of \mbox{Her X-1} during the dip is due to absorption and scattering in the
intervening material. The absolute value of $I_\mathrm{U}$ is quite
small (about 2.5\% of $I_\mathrm{A}$). 

The normalization of the iron line, $N_{\rm Fe}$, also shows some decline
during phases of high column density. Note, however, that in our spectral
model the line feature is not absorbed and scattered. The remaining
variation can in principle be explained by partial covering of the line
emitting region.

\begin{figure*}
  \resizebox{12cm}{!}{\includegraphics{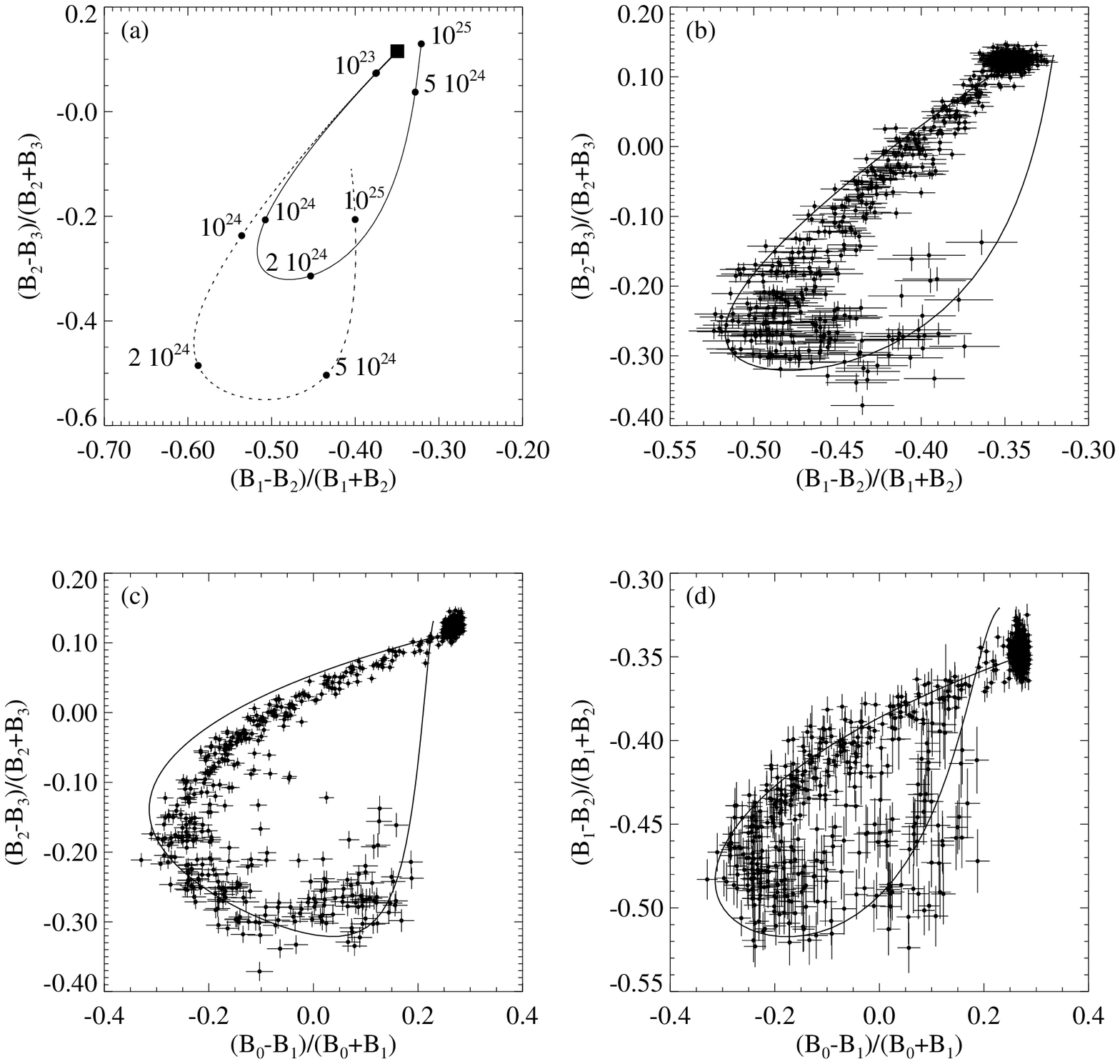}}
\hfill%
\parbox[b]{55mm}{%
  \caption{\textbf{a} Theoretical track of a partial covering model (solid
    line) and a model without unabsorbed component (dashed line) in a color
    color diagram for spectral parameters typical for Her~X-1. Labels on
    the curve refer to column densities (in $\mbox{cm}^{-2}$).
    \textbf{b}--\textbf{d} Color-color diagrams and best fit partial
    covering models for the dip of the RXTE observation.}\label{fig:ccd} }
\end{figure*}

\begin{figure*}
  \resizebox{\hsize}{!}{\includegraphics{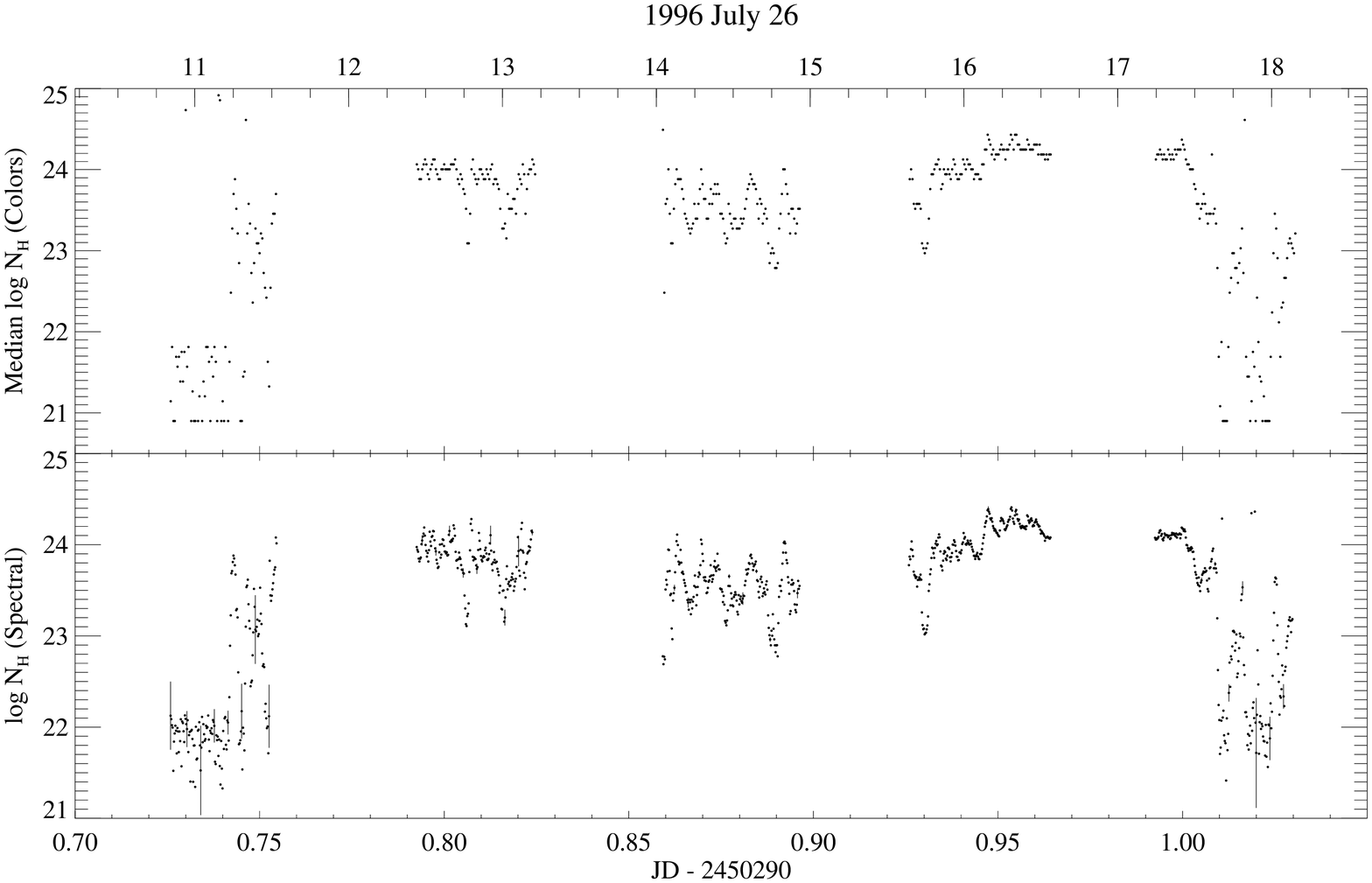}}
  \caption{Comparison of the time development of $N_\mathrm{H}$ as found from 
    different methods: Upper panel: $N_\mathrm{H}$ from color-color
    diagrams, lower panel: $N_\mathrm{H}$ from spectral fitting.}
\label{fig:nh}
\end{figure*}

\subsection{Color-color diagrams}\label{subsec:colordiag}
As an alternative to spectral fitting, the development of the column
density can be visualized by color-color diagrams which show the behavior
of broad band X-ray count rates
(cf. \cite{Leahy95.2} for earlier results from \textsl{Ginga} data). We
define four energy bands covering
approximately the same range as our spectral analysis
(Tab.~\ref{tab:energybands}). We then define modified X-ray colors by
$(B_0-B_1)/(B_0+B_1)$, $(B_1-B_2)/(B_1+B_2)$, and $(B_2-B_3)/(B_2+B_3)$
where $B_i$ is the count rate in band $i$.
\begin{table}
  \caption{Energy bands defined for the analysis of
    colors.}\label{tab:energybands} 
  \begin{center}
    \begin{tabular}{lll}\hline
      Band &  Energy [keV] & PHA Channels \\ \hline
      $B_0$ & ~3--~5 & ~4--11 \\
      $B_1$ & ~5--~7 & 12--15 \\
      $B_2$ & ~7--11 & 16--26 \\
      $B_3$ & 11--17 & 27--42 \\ \hline
    \end{tabular}
  \end{center}
\end{table}
For any given spectral model, a theoretical color can be obtained by
folding the spectral model through the detector response matrix. If the
only variable parameter in the model is $N_\mathrm{H}$, the resulting 
colors are found to trace characteristic tracks in the color-color diagram.
Comparing these tracks with the measured data it is possible to infer the
temporal behavior of the column density (see below).

As an example, Fig.~\ref{fig:ccd}a displays typical theoretical tracks for
two possible spectral models for Her~X-1. The dotted track represents a
model without an unabsorbed component (called the one-component model
henceforth), while the solid line is the track computed for a partial
covering model. The form of this partial covering model is identical to
Eq.~(\ref{eq:pcovmod}) with the exception that the iron line is also
absorbed and scattered. We used photoabsorption cross sections from
\citey{Verner95.1} and \citey{Verner96.1} in the computation of the
diagrams. The difference between these cross sections and those from
\citey{Morrison83.1} used in Sect.~\ref{subsec:specfits} is negligible,
though.  The typical shape of the tracks is due to the $E^{-3}$
proportionality of the absorption cross section $\sigma_\mathrm{bf}$: for
low values of $N_\mathrm{H}$ only the lower bands are influenced by the
absorbing material, while for high values of $N_\mathrm{H}$ all bands are
influenced.  In both cases the model track starts at the low
$N_\mathrm{H}$-values in the upper right corner of the diagram marked by
the square which describes the situation before the dip.  Moving along the
track the column density increases. For low values of $N_\mathrm{H}$, the
tracks of both models are similar since the influence of absorption is
negligible. For larger $N_\mathrm{H}$ the lower energy bands are
increasingly affected by absorption. At a critical value of $N_\mathrm{H}$
the unabsorbed component begins to dominate the low energy bands in the
partial covering model. Since the unabsorbed component has, by definition,
the same shape as the non-dip spectrum, the track turns towards the
low-$N_\mathrm{H}$ color. In the one-component model, the absence of an
unabsorbed component leads to a further decrease in flux in the low energy
bands that is only stopped by response matrix and detector background
effects.

For each of the energy bands defined in Table~\ref{tab:energybands} we
generate a background subtracted lightcurve of 32\,s resolution and obtain
the color-color diagrams shown in Fig.~\ref{fig:ccd}b--d.  The data line
up along a track which is curved similar to the theoretical tracks of
Fig.~\ref{fig:ccd}a.  The accumulation of data points in the upper right
corner of the diagrams of Fig.~\ref{fig:ccd}b--d consists of the
out-of-dip data, where the colors remain constant and the column density is
at its lowest value.  As noted above, the early turn of the observed tracks
in the color-color diagram suggests the presence of an unabsorbed spectral
component in the data.  To quantify this claim, we compare theoretical
tracks from the partial covering model to the data. This is done by varying
the relative contributions of $I_\mathrm{A}$ and $I_\mathrm{U}$ to the
total spectrum such that the total normalization of the incident spectrum,
$I_\mathrm{A}+I_\mathrm{U}$, is kept at its pre-dip value. Except for the
normalizations, all other spectral parameters are fixed at their pre-dip
values (cf.\ Sect.~\ref{subsec:specfits}).  We define the best fit model to
be the model in which the root mean square distance between the track and
the data is minimal (cf.\ Fig.~\ref{fig:ccd}).  Not surprisingly, the ratio
between $I_\mathrm{A}$ and $I_\mathrm{U}$ found using this method is
similar to the average ratio found from spectral fitting, about 3\%.

Using this best fit model, $N_\mathrm{H}$ as a function of time is found by
projecting the measured colors onto the track.  The projection provides
slightly different values of $N_\mathrm{H}$ for each color-color diagram 
examined.  Major discrepancies are due to projection onto a wrong part of
the model curve in the region where the curve overlaps with itself. To even
out these discrepancies we calculated the median of $N_\mathrm{H}$ from all
three color-color diagrams used in our analysis.  The time development of
$N_\mathrm{H}$ found from the color-color diagrams is in good agreement
with the $N_\mathrm{H}$ resulting from the spectral fits
(Fig.~\ref{fig:nh}).

\section{Light curves}\label{sec:lightcurves}
The irregular variations observed in the lightcurve during the dips are
generally thought to be due to the cloudy structure of an absorber
extending above the disk surface.  The location of the material, i.e., its
distance from the neutron star, however, is unknown and depends on model
assumptions. According to \citey{Crosa80}, the obscuring matter is the
temporarily thickened disk rim, or, in the modification of this model by
\citey{Bochkarev89.1} and \citey{Bochkarev89.2}, consists of ``blobs'' in
an extended corona above the disk rim.  On the other hand, in the coronal
wind model of \citey{Schandl97.1} the dips are caused by a spray of matter
at some inner disk radius where the accretion stream impacts on the disk.

\begin{figure*}
  \resizebox{\hsize}{!}{\includegraphics{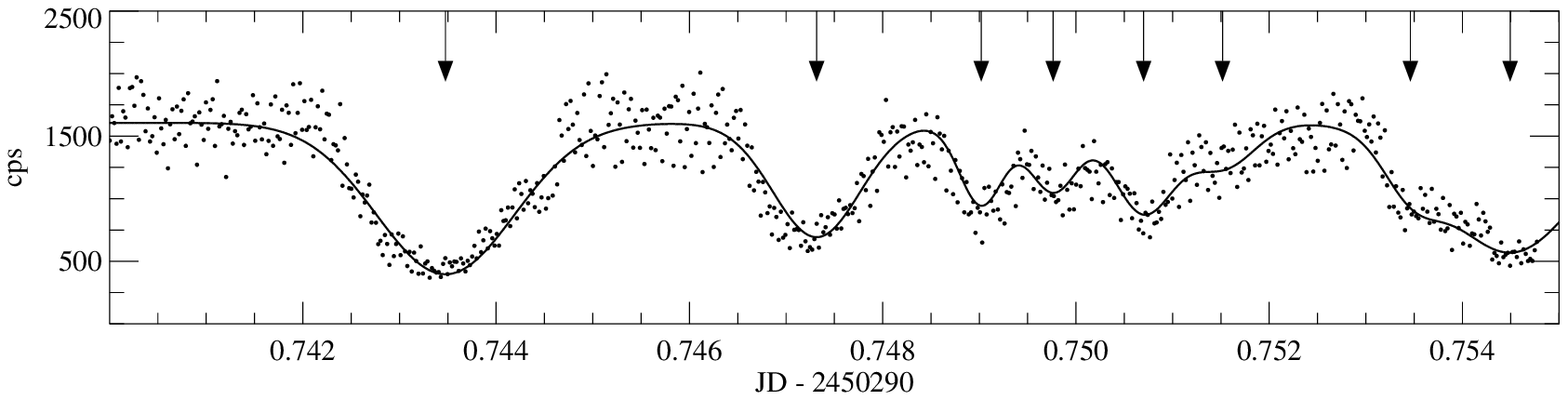}}
  \resizebox{\hsize}{!}{\includegraphics{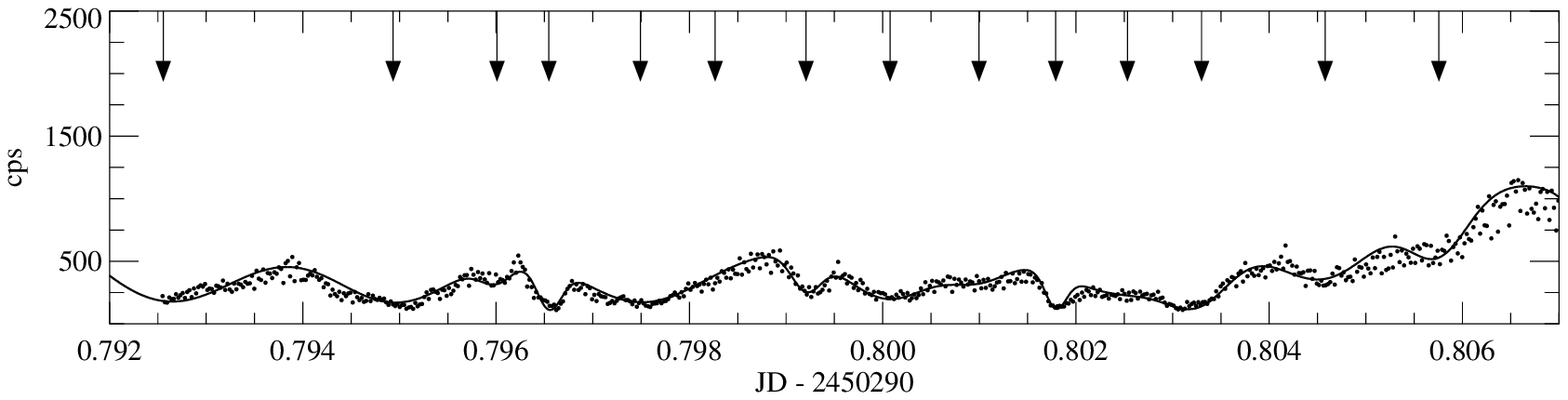}}
  \resizebox{\hsize}{!}{\includegraphics{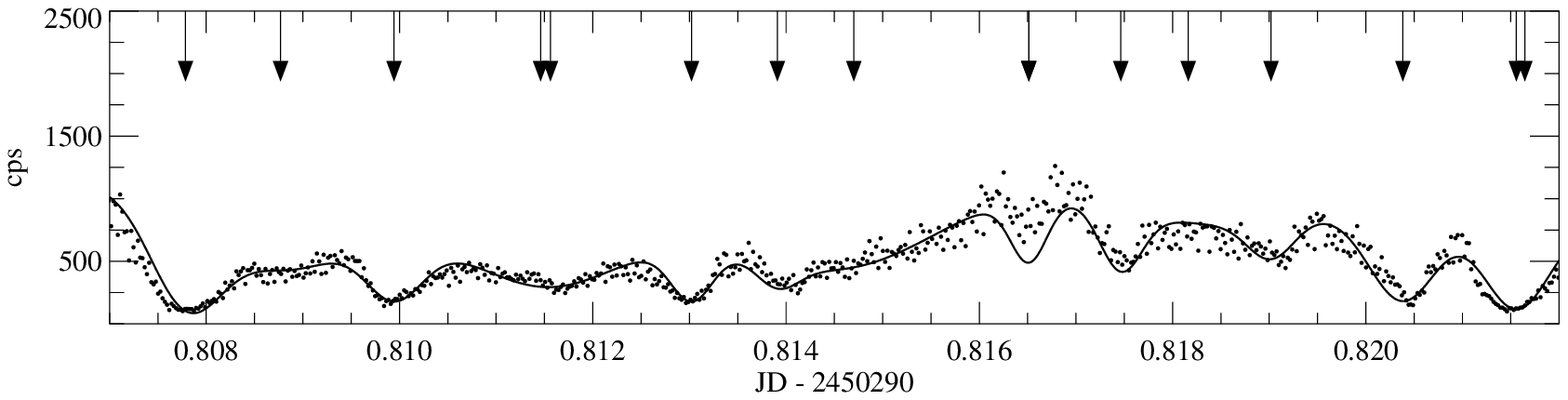}}
\caption{Gaussian fits to the dip lightcurve with a temporal resolution of
  2\,s. Arrows mark the position of Gaussian line centroids.}
\label{fig:lcfit}
\end{figure*}

Since all models assume that the absorbing matter exists in the form of
clumps of material, it is interesting to ask whether these blobs can be
seen in the data. The symmetry of the structures in the first half of the
lightcurve shown in Fig.~\ref{fig:timeevol}a suggests that this is indeed the
case.  We therefore tried to model this part of the dip lightcurve by
fitting Gaussians to each of the visibly identified structures.  The best
fit to the data is displayed in Fig.~\ref{fig:lcfit}.  Information about
the parameters of each Gaussian is summarized in Table~2 
(available in electronic form only). Note
that the structures appear to be quasi-periodic on a time-scale of about
$10^{-3}$\,d (=90\,s) which could provide a hint on their nature. 
A similar 144\,s periodicity has previously been reported by
\citey{Leahy92.1}.  A study of a larger sample of dips is necessary, however,
before any claim on the existence of periodic structures in the dip data
can be settled.

Assuming that each structure in the lightcurve represents a single cloud of
matter which is absorbing X-rays while crossing the line of sight, the
width of the Gaussian represents a direct measure of the crossing time for
such a cloud.  The horizontal extent of the cloud could be derived if the
radial distance of these clouds from the neutron star and their velocity
were known.  To give a rough estimate for the cloud size, we assume that
the matter moves on Keplerian orbits close to the outer disk rim (as is
favored by the model of \cite{Crosa80}), at a distance of $2\,10^{11}\,$cm
from the neutron star (\cite{Cheng95}).  The Keplerian velocity at this
radius is approximately $3\,10^7$\,cm/s. Assuming that the clouds are
spherical, i.e., assuming that their angular size estimated from the FWHM
of the fitted Gaussian corresponds to their radial size, we derive typical
cloud diameters on the order of $10^{8\ldots 10}$\,cm. Making use of the
observed column densities we find a proton density of about
$6\,10^{14}\,\mbox{cm}^{-3}$. On the other hand, the apparent periodicity
mentioned above might also indicate that the blobs are very close to the
neutron star. Assuming Keplerian motion, the periodicity could indicate
orbital radii as small as $10^9$\,cm (the inner radius of the accretion
disk is at $10^8$\,cm; \cite{Horn92.1}) and their densities would be
accordingly higher.

\section{Discussion and Conclusions}\label{sec:conclusions}
In this paper we have presented the temporal evolution of spectral
parameters of \mbox{Her X-1} during a pre-eclipse dip. Due to the large
effective area of the PCA, the temporal resolution of our data is higher
than that of earlier data.  Using two different methods we were able to
show that a partial covering model in which the column density
$N_\mathrm{H}$ and the normalization of the unabsorbed continuum
$I_\mathrm{U}$ are the only time variable spectral components is sufficient
to explain the observed spectral and temporal variability.  This result is
in qualitative agreement with the previous analyses for pre-eclipse dips
presented by \citey{Ushimaru89}, \citey{Choi94.1}, \citey{Leahy94.1}, and
\citey{Reynolds95.1}.  Note that models without an unabsorbed component
were not able to result in satisfactory fits, which is also consistent with
the low state observations presented by \citey{Mihara91.1} where the
unabsorbed component was clearly required. As proposed by \citey{Choi94.1}
from the analysis of the pulsed fraction of the lightcurve, the unabsorbed
component is probably due to scattering of radiation in an extended hot
electron corona into the line of sight. These authors also show that the
interpretation of \citey{Ushimaru89}, who attributed the unabsorbed
component to a leaky cold absorber, does not hold.

The anticorrelation between $I_\mathrm{U}$ and $N_\mathrm{H}$ found by
\citey{Leahy94.1} and \citey{Leahy97.1} in \textsl{Ginga} observations has
been interpreted by these authors as evidence that the obscuring material
partially also obscures the extended corona. Thus, the geometric covering
fraction of the obscuring material is assumed to be quite high during
episodes of large $N_\mathrm{H}$. Our measured values of $I_\mathrm{U}$
exhibit some variability (Fig.~\ref{fig:timeevol}c), but there is no
systematic correlation between $I_\mathrm{U}$ and $N_\mathrm{H}$, neither
has such a correlation been seen by \citey{Choi94.1}. A possible
interpretation for this discrepancy is that \citey{Leahy94.1} and
\citey{Leahy97.1} did not include Thomson scattering in their fits. Indeed,
when setting $k=0$ in Eq.~(\ref{eq:pcovmod}) and fitting the RXTE data with
both normalizations as free parameters, $I_\mathrm{U}$ is much more
variable and $I_\mathrm{A}$ appears to be correlated with $N_\mathrm{H}$.
In addition, it is generally difficult to distinguish between the absorbed
and the unabsorbed component for low values of $N_\mathrm{H}$ such that
$I_\mathrm{U}$ and $I_\mathrm{A}$ get easily confused by the fitting
routine.

We also tried fitting the data with a spectral model in which 
$I_\mathrm{U}$ was held fixed at \linebreak
$0.005\,\mbox{ph}\,\mbox{cm}^{-2}\,\mbox{s}^{-1}\,\mbox{keV}^{-1}$, the
average value of the fits of Sect.~\ref{subsec:specfits}. The resulting
$\chi^2_\mathrm{red}$ values from this fit were comparable to those of the
fits presented in Sect.~\ref{subsec:specfits} since the variations in
$I_\mathrm{U}$ in the latter fit are small enough to be compensated by
slight changes in $N_\mathrm{H}$ in the former.  Furthermore, due to the
lower temporal resolution of the previous observations, small variations of
$N_\mathrm{H}$ could not be resolved in these data.  It has been pointed
out by \citey{Parmar86.1}, that this effect might result in a large
uncertainty in the determination of $I_\mathrm{U}$. As is shown by our fits
to structures in the lightcurve (Fig.~\ref{fig:lcfit}), we are able to
resolve and identify individual structures with a temporal resolution of
about one minute. Thus we are confident that the investigation presented
here is unaffected by these problems.

In our analysis of Sect.~\ref{subsec:specfits} and \ref{subsec:colordiag}
we assumed that $N_\mathrm{e}/N_\mathrm{H}=1.21$, i.e., the value
appropriate for material of solar composition.  Previous investigations of
dipping sources, however, hinted at non-solar abundances in most of these
systems (\cite{Reynolds95.1}, \cite{White}).  A direct measurement of the
abundance ought to be possible by fitting the RXTE data with a partial
covering model in which $N_\mathrm{e}$ and $N_\mathrm{H}$ are both free
parameters. We find that the average $N_\mathrm{e}/N_\mathrm{H}=1.5$.
Thus, our data could be interpreted as pointing towards a metal
overabundance.  In contrast to this result, a metal underabundance is
favored by \citey{Reynolds95.1}\footnote{Note that \citey{Reynolds95.1}
  assume that for material of cosmic abundance
  $N_\mathrm{e}/N_\mathrm{H}=1$, and call the reciprocal of the measured
  ratio the ``abundance''. This reciprocal should not be confused with the
  common usage of abundance in astronomy since it ignores that the
  scattering cross section of an element with nuclear charge number $Z$ is
  $Z\sigma_\mathrm{T}$.}. This discrepancy might be due to the higher
temporal resolution of our data, where small variations of $N_\mathrm{e}$
can be traced. Note, however, that our method of determining
$N_\mathrm{e}/N_\mathrm{H}$ depends crucially on the assumption that the
spectrum incident on the absorbing cloud is constant over time and also
depends strongly on the value adopted for $I_\mathrm{A}$. Although it is
very probable that the source is constant (Sect.~\ref{subsec:specfits}),
slight variations of $I_\mathrm{A}$ cannot be ruled out.  Therefore, we
decided to use the solar value of 1.21 in our analysis and postpone the
detailed study of the abundances until a larger sample of dips has been
observed with RXTE.

To conclude, photoabsorption and electron scattering of photons out of the
line of sight in cold material appear to be solely sufficient for
explaining the temporal variability of the observed flux during this
pre-eclipse dip of \mbox{Her X-1}. Further observations of a larger sample
of pre-eclipse dips are necessary to verify this result.

\begin{acknowledgements}
  We acknowledge W.~Heindl, I.~Kreykenbohm, and R.~Meier for useful
  discussions on the data analysis. We thank G.~Morfill and R.~Neuh\"auser
  for granting the first author the time to finish this work.  This work
  has been financed by DARA grant 50\,OR\,92054.
\end{acknowledgements}


\begin{thebibliography}{}
  
\bibitem[\protect\astroncite{Arnaud}{1996}]{Arnaud96.1} Arnaud K.A., 1996,
\newblock In: Jacoby J.H., Barnes J. (eds.) Astronomical Data Analysis
  Software and Systems {V}. Astron.\ Soc.\ Pacific, Conf.\ Ser. 101,
  Astron.\ Soc.\ Pacific, San Francisco, p.~17

\bibitem[\protect\astroncite{Bochkarev}{1989}]{Bochkarev89.1}
Bochkarev N.G.,  1989, SvA 33, 638

\bibitem[\protect\astroncite{Bochkarev \& Karitskaya}{1989}]{Bochkarev89.2}
Bochkarev N.G., Karitskaya E.A.,  1989, Ap\&SS 154, 189

\bibitem[\protect\astroncite{Cheng et~al.}{1995}]{Cheng95}
Cheng F.H., Vrtilek S.D., Raymond J.C.,  1995, ApJ 452, 825

\bibitem[\protect\astroncite{Choi et~al.}{1994}]{Choi94.1}
Choi C.S., Nagase F., Makino F., et~al., 1994, ApJ 422, 799

\bibitem[\protect\astroncite{Crosa \& Boynton}{1980}]{Crosa80}
Crosa L., Boynton P.E.,  1980, ApJ 235, 999

\bibitem[\protect\astroncite{Dove et~al.}{1998}]{Dove98.1}
Dove J.B., Wilms J., Nowak M.A., et~al., 1998, MNRAS 298, 729

\bibitem[\protect\astroncite{Giacconi et~al.}{1973}]{Giacconi73.1}
Giacconi R., Gursky H., Kellogg E., et~al., 1973, ApJ 184, 227

\bibitem[\protect\astroncite{Horn}{1992}]{Horn92.1}
Horn S.,  1992,
\newblock Dissertation, Ludwig-Maximilians-Universit\"at M\"unchen

\bibitem[\protect\astroncite{Jahoda}{1996}]{Jahoda96.2}
Jahoda K.,  1996,
\newblock Estimating the {B}ackground in the {PCA},
\newblock Technical report, NASA GSFC, Greenbelt version dated November 27,
  1996

\bibitem[\protect\astroncite{Jahoda et~al.}{1997}]{Jahoda96.1}
Jahoda K., Swank J.H., Giles A.B., et~al., 1997,
\newblock In: Siegmund O.H. (ed.) {EUV}, X-Ray, and Gamma-Ray Instrumentation
  for Astronomy {VII}. Proc.\ SPIE 2808, SPIE, Bellingham, WA, p.59

\bibitem[\protect\astroncite{Leahy}{1995}]{Leahy95.2}
Leahy D.A.,  1995, A\&AS 113, 21

\bibitem[\protect\astroncite{Leahy}{1997}]{Leahy97.1}
Leahy D.A.,  1997, MNRAS 287, 622

\bibitem[\protect\astroncite{Leahy et~al.}{1992}]{Leahy92.1}
Leahy D.A., Yoshida A., Kawai N., Matsuoka M.,  1992,
\newblock In: Shrader C.R., Gehrels N., Dennis B. (eds.) The Compton
  Observatory Science Workshop., NASA CP 3137, p.~193

\bibitem[\protect\astroncite{Leahy et~al.}{1994}]{Leahy94.1}
Leahy D.A., Yoshida A., Matsuoka M.,  1994, ApJ 434, 341

\bibitem[\protect\astroncite{Mavromatakis}{1993}]{Mavromatakis93}
Mavromatakis F.,  1993, A\&A 273, 147

\bibitem[\protect\astroncite{Mihara et~al.}{1991}]{Mihara91.1}
Mihara T., Ohashi T., Makishima K.,  1991, PASJ 43, 501

\bibitem[\protect\astroncite{Morrison \& McCammon}{1983}]{Morrison83.1}
Morrison R., McCammon D.,  1983, ApJ 270, 119

\bibitem[\protect\astroncite{Parmar et~al.}{1986}]{Parmar86.1}
Parmar A.N., White N.E., Giommi P., Gottwald M.,  1986, ApJ 308, 199

\bibitem[\protect\astroncite{Reynolds \& Parmar}{1995}]{Reynolds95.1}
Reynolds A.P., Parmar A.N.,  1995, A\&A 297, 747

\bibitem[\protect\astroncite{Schandl}{1996}]{Schandl96.1}
Schandl S.,  1996, A\&A 307, 95

\bibitem[\protect\astroncite{Schandl et~al.}{1997}]{Schandl97.1} 
Schandl S., K\"onig M., Staubert R., 1997, \newblock In: Dermer C.D.,
Strickman M.S., Kurfess J.D. (eds.) Proc.\ 4th Compton Symposium, AIP
Conf.\ Proc.\ 410, Woodbury: AIP, p.763

\bibitem[\protect\astroncite{Stelzer et~al.}{1997}]{Stelzer97.1}
Stelzer B., Staubert R., Wilms J., Geckeler R.D.,  1997,
\newblock In: Dermer C.D., Strickman M.S., Kurfess J.D. (eds.) Proc.\
4th Compton Symposium, AIP Conf.\ Proc.\ 410, Woodbury: AIP, p.~753

\bibitem[\protect\astroncite{Ushimaru et~al.}{1989}]{Ushimaru89}
Ushimaru N., Tawara Y., Koyama K., Hayakawa S.,  1989, PASJ 41, 441

\bibitem[\protect\astroncite{Verner \& Yakovlev}{1995}]{Verner95.1}
Verner D.A., Yakovlev D.G.,  1995, A\&AS 109, 125

\bibitem[\protect\astroncite{Verner et~al.}{1996}]{Verner96.1}
Verner D.A., Ferland G.J., Korista K.T., Yakovlev D.G.,  1996, ApJ 465, 487

\bibitem[\protect\astroncite{White et~al.}{1995}]{White}
White N.E., Nagase F., Parmar A.N.,  1995, 
\newblock In: Lewin W. H. G., van Paradijs J., van den Heuvel E. P. J. (eds.)
X-Ray Binaries, Cambridge Univ.\ Press, Cambridge, p.1

\end{thebibliography}
\end{document}